\documentclass[10pt,nofootinbib,twocolumn,aps,prd]{revtex4-1}
\usepackage{amsmath,mathptmx,amssymb}
\usepackage{graphicx,graphics,subfigure}
\usepackage{tabularx,ctable}%,multirow}
\usepackage[colorlinks=true,linkcolor=blue,citecolor=blue,urlcolor=blue]{hyperref}

\def\be{\begin{equation}}
\def\ee{\end{equation}}
\def\bea{\begin{eqnarray}}
\def\eea{\end{eqnarray}}
\def\ba{\begin{aligned}}
\def\ea{\end{aligned}}
\def\nn{\nonumber}
\def\p{\partial}
\def\mbA{\mathbb{A}}
\def\cN{\mathcal{N}}
\def\cF{\mathcal{F}}

\begin{document}

\title{Consistent thermodynamics and topological classes for the four-dimensional Lorentzian
charged Taub-NUT spacetimes}

\author{Di Wu}
\email{wdcwnu@163.com}
% https://orcid.org/0000-0002-2509-6729

\affiliation{School of Physics and Astronomy, China West Normal University,
Nanchong, Sichuan 637002, People's Republic of China}

\date{\today}

\begin{abstract}
In this paper, we derive the consistent thermodynamics of the four-dimensional Lorentzian
Reissner-Nordstr\"om-NUT (RN-NUT), Kerr-Newman-NUT (KN-NUT), and RN-NUT-AdS spacetimes in
the framework of the ($\psi-\cN$)-pair formalism, and then investigate their topological
numbers by using the uniformly modified form of the generalized off-shell Helmholtz free
energy. We find that these solutions can be included into one of three categories of those
well-known black hole solutions, which implies that these spacetimes should be viewed as
generic black holes from the perspective of the topological thermodynamic defects. In
addition, we demonstrate that although the existence of the NUT charge parameter seems
to have no impact on the topological number of the charged asymptotically locally flat
spacetimes, it has a remarkable effect on the topological number of the charged asymptotically
locally AdS spacetime.
\end{abstract}

\maketitle

%\tableofcontents

%%%%%%%%%%%%%%%%%%%%%%
\section{Introduction}
%%%%%%%%%%%%%%%%%%%%%%

In recent years, the study of the topology of black holes has attracted much interest and attention,
including light rings \cite{PRL119-251102,PRL124-181101,PRD102-064039,PRD103-104031,PRD105-024049},
timelike circular orbits \cite{PRD107-064006,2301.04786}, thermodynamics \cite{PRD105-104003,
PRD105-104053,PLB835-137591,PRD107-046013,PRD107-106009,JHEP0623115,2305.05595,2305.05916,2305.15674,
2305.15910}, phase transitions \cite{PRD106-064059,PRD107-044026,PRD107-064015,2212.04341,2302.06201,
2304.14988}, as well as thermodynamic topological classifications \cite{PRL129-191101,PRD107-064023,
JHEP0123102,PRD107-024024,PRD107-084002,PRD107-084053,2303.06814,2303.13105,2304.02889,2306.13286,
2304.05695,2306.05692,2306.11212,EPJC83-365}, and so on. \textcolor{black}{In particular, a novel approach
has been recently suggested in Ref. \cite{PRL129-191101} to study} the thermodynamic topological properties
of black holes by interpreting black hole solutions as topological thermodynamic defects, creating
topological numbers, and then classifying all black holes into three separate categories based upon their
different topological numbers, which sheds new light on the fundamental properties of black holes
and gravity. Due to its adaptability and simplicity, the topological approach suggested in Ref.
\cite{PRL129-191101} quickly gained widespread acceptance. As a result, it was successfully applied to
explore the topological numbers of several well-known black hole solutions \cite{PRD107-064023,JHEP0123102,
PRD107-024024,PRD107-084002,PRD107-084053,2303.06814,2303.13105,2304.02889,2306.13286,2304.05695,2306.05692,
2306.11212}, for examples, the static Gauss-Bonnet-AdS black holes \cite{PRD107-064023}, the static black
hole in nonlinear electrodynamics \cite{JHEP0123102}, the Kerr and Kerr-Newman black holes
\cite{PRD107-024024}, the Kerr-AdS and Kerr-Newman-AdS as well as three-dimensional BTZ black holes
\cite{PRD107-084002}, some static hairy black holes \cite{PRD107-084053}, the dyonic black hole in
nonlinear electrodynamics \cite{2303.06814}, the black hole in de Sitter spacetimes \cite{2303.13105}, and
the black hole in massive gravity \cite{2304.02889,2306.13286}, the static dyonic AdS black holes in
different ensembles \cite{2304.05695}, \textcolor{black}{some Bardeen black holes \cite{2306.05692}, as well
as the static Born-Infeld-AdS black holes \cite{2306.11212}}. Very recently, we have investigated the
topological numbers for the cases of the four-dimensional Lorentzian Taub-NUT, Kerr-NUT, and Taub-NUT-AdS$_4$
spacetimes \cite{EPJC83-365}, and demonstrated that these spacetimes should be viewed as generic black holes
from the viewpoint of the thermodynamic topological approach. It is then natural for us to extend that work
to the more general charged cases with a pure electric charge to examine whether the four-dimensional
Lorentzian charged Taub-NUT spacetimes are generic black holes, which serves as our motivation for the
present work.

In this paper, \textcolor{black}{within the framework of the ($\psi-\cN$)-pair formalism, we will
first utilize the generalized Komar super-potential \cite{JHEP1022174} to} derive the consistent
thermodynamics of the four-dimensional Lorentzian Reissner-Nordstr\"om-NUT (RN-NUT), Kerr-Newman-NUT
(KN-NUT), and RN-NUT-AdS spacetimes, and to investigate their topological number via the uniformly
modified form of the generalized off-shell Helmholtz free energy. We find that these spacetimes
should also be viewed as generic black holes from the thermodynamic topological perspective.

The remaining part of this paper is organized as follows. In Sec. \ref{II}, we give a brief review
of the novel thermodynamic topological approach for Taub-NUT-type spacetimes. In Sec. \ref{III}, we
first derive the consistent formulation of thermodynamic properties of the four-dimensional Lorentzian
RN-NUT spacetime and then investigate its topological number. In Sec. \ref{IV}, we turn to discuss
the case of the Lorentzian KN-NUT spacetime. In Sec. \ref{V}, we then extend to discuss the more
general Lorentzian RN-NUT-AdS$_4$ spacetime. Finally, we present our conclusions in Sec. \ref{VI}.

%%%%%%%%%%%%%%%%%%%%%%%%%%%%%%%%%%%%%%%%%%%
\section{Thermodynamic topological approach
for Taub-NUT-type spacetimes}\label{II}
%%%%%%%%%%%%%%%%%%%%%%%%%%%%%%%%%%%%%%%%%%%

In accordance with the thermodynamic topological approach suggested in Ref. \cite{EPJC83-365}, it
is possible to introduce the modified form of the generalized off-shell Helmholtz free energy
\be\label{FE}
\cF = M -\frac{S}{\tau} -\psi\cN \, ,
\ee
for a Taub-NUT-type black hole thermodynamical system with the mass $M$, the entropy $S$, and
the Misner potential $\psi$, as well as the gravitational Misner charge $\cN$, where $\tau$ is
an extra variable that can be treated as the inverse temperature of the cavity surrounding the
Taub-NUT-type black hole. Only when $\tau = T^{-1}$, the modified form of the generalized Helmholtz
free energy (\ref{FE}) is on-shell and reduces to the common Helmholtz free energy: $F = M -TS
-\psi\cN$ of the Taub-NUT-type black holes \cite{PRD100-064055,CQG36-194001,JHEP0520084}.

According to Ref. \cite{PRL129-191101}, a core vector $\phi$ is defined as
\bea\label{vector}
\phi = \Big(\frac{\p \cF}{\p r_{h}}\, , ~ -\cot\Theta\csc\Theta\Big) \, ,
\eea
where the two parameters satisfy the ranges: $0 < r_h < +\infty$, $0 \le \Theta \le \pi$, respectively.
The component $\phi^\Theta$ is divergent at $\Theta = 0$ and $\Theta = \pi$, indicating that the
direction of the vector is outward there.

One can define the topological current by using Duan's $\phi$-mapping topological current theory
\cite{SS9-1072,NPB514-705,PRD61-045004} as follows:
\be\label{jmu}
j^{\mu}=\frac{1}{2\pi}\epsilon^{\mu\nu\rho}\epsilon_{ab}\p_{\nu}n^{a}\p_{\rho}n^{b}\, , \qquad
\mu,\nu,\rho=0,1,2,
\ee
where $\p_{\nu}= \p/\p x^{\nu}$ and $x^{\nu}=(\tau,~r_h,~\Theta)$. The unit vector $n$ is $n =
(n^r, n^\Theta)$, where $n^r = \phi^{r_h}/||\phi||$ and $n^\Theta = \phi^{\Theta}/||\phi||$. Since
it is easy to prove that the above current (\ref{jmu}) is conserved, and one can quickly obtain
$\p_{\mu}j^{\mu} = 0$ and then indicate that the topological current is a $\delta$-function of
the field configuration \cite{NPB514-705,PRD61-045004}
\be
j^{\mu}=\delta^{2}(\phi)J^{\mu}\Big(\frac{\phi}{x}\Big) \, ,
\ee
where the three dimensional Jacobian $J^{\mu}(\phi/x)$ obeys: $\epsilon^{ab}J^{\mu}(\phi/x) =
\epsilon^{\mu\nu\rho}\p_{\nu}\phi^a\p_{\rho}\phi^b$. It is easy to demonstrate that $j^\mu$ equals
to zero only when $\phi^a(x_i) = 0$, and one can easily obtain the topological number $W$ as follows:
\be
W = \int_{\Sigma}j^{0}d^2x = \sum_{i=1}^{N}\beta_{i}\eta_{i} = \sum_{i=1}^{N}w_{i}\, ,
\ee
where $\beta_i$ is the positive Hopf index that counts the number of the loops of the vector $\phi^a$
in the $\phi$-space when $x^{\mu}$ are around the zero point $z_i$, while $\eta_{i}= \mathrm{sign}
(J^{0}({\phi}/{x})_{z_i})=\pm 1$ is the Brouwer degree, and $w_{i}$ is the winding number for the
$i$-th zero point of $\phi$ that is contained in the domain $\Sigma$. \textcolor{black}{In addition,
if two different closed curves $\Sigma_1$ and $\Sigma_2$ enclose the same zero point of $\phi$, the
corresponding winding number must equal. On the other hand, if there is no zero point of $\phi$ in
the enclosed region, one must have $W = 0$.}

\textcolor{black}{Note that the local winding number $w_{i}$ can be used to characterize the local
thermodynamic stability, with positive and negative values corresponding to thermodynamically
stable and unstable black holes, respectively, and the global topological number $W$ represents
the difference between the number of thermodynamically stable black holes and the number of
thermodynamically unstable black holes of a classical black hole solution at a fixed temperature
\cite{PRL129-191101}. Therefore, not only can one distinguish between different black hole phases
(thermodynamically stable or unstable) of the same black hole solution at a specific temperature
based upon the local winding number, but also can classify the black hole solutions based upon
the global topological number. Furthermore, according to this classification, black holes with
the same global topological number (even if they are of different geometric types) have similar
thermodynamical topology properties.}

%%%%%%%%%%%%%%%%%%%%%%%%%%%%%%%%%%%%%
\section{RN-NUT spacetime}\label{III}
%%%%%%%%%%%%%%%%%%%%%%%%%%%%%%%%%%%%%

As the simplest charged case, we will investigate the four-dimensional Lorentzian RN-NUT solution
\cite{PR133-B845,PRD100-101501,2210.17504}, and adopt the following line element in which the Misner
strings are symmetrically distributed along the polar axis:
\bea\label{RNNUT}
ds^2 &=& -\frac{f(r)}{r^2 +n^2}(dt +2n\cos\theta\, d\varphi)^2 +\frac{r^2 +n^2}{f(r)}dr^2 \nn \\
&& +\big(r^2 +n^2\big)\big(d\theta^2 +\sin^2\theta\, d\varphi^2\big) \, ,
\eea
where
\be
f(r) = r^2 -2mr -n^2 +q^2 \, , \nn
\ee
in which $m$, $n$ and $q$ are the mass, the NUT charge, and the electric charge parameters,
respectively. The event horizon radius: $r_h = m +\sqrt{m^2 +n^2 -q^2}$ is the largest root
of the equation: $f(r_h) = 0$. In addition, the electromagnetic gauge potential one-form is
given by
\be\label{Abel}
\mbA = \frac{qr}{r^2+n^2}(dt +2n\cos\theta\, d\varphi) \, ,
\ee
\textcolor{black}{with which a convenient gauge choice is made so that its temporal component
vanishes at infinity}.

\textcolor{black}{The metric (\ref{RNNUT}) with the Abelian gauge potential (\ref{Abel}) is
an exact solution to the field equations derived from the Lagrangian density: $\mathcal{L} =
\sqrt{-g}\big(R -F^2\big)/(16\pi)$. For latter convenience, we introduce a generalized Komar
super-potential (see Eq. (5.20) of Ref. \cite{JHEP1022174}) as follows:
\be
\Xi^{ab}[\xi] = \nabla^a\xi^b -\nabla^b\xi^a +(4\xi^c{}F^{ab} +2\xi^a{}F^{bc}
 +2\xi^b{}F^{ca})\mbA_c \, , \label{gKsp}
\ee
associated with a Killing vector $\xi$ and the Faraday-Maxwell field strength tensor:
$F_{ab} = \nabla_a\mbA_b -\nabla_b\mbA_a$ defined by $F = d\mbA$. It can be shown that
$\nabla_b\Xi^{ab} = 0$.}

%%%%%%%%%%%%%%%%%%%%%%%%%%%%%%%%%%%%%%%%%%%%%%%%%%
\subsection{Consistent thermodynamics}\label{IIIA}
%%%%%%%%%%%%%%%%%%%%%%%%%%%%%%%%%%%%%%%%%%%%%%%%%%

As \textcolor{black}{we are only focused on the purely electrically charged RN-NUT solutions,
so} we will first rederive the consistent thermodynamics of the four-dimensional Lorentzian
RN-NUT spacetime within the framework of the ($\psi-\cN$)-pair formalism.

The Bekenstein-Hawking entropy is one quarter of the area of the event horizon
\be
S = \frac{A}{4} = \pi\big(r_h^2 +n^2\big) \, ,
\ee
the Gibbons-Hawking temperature is proportional to the surface gravity $\kappa$ on the event
horizon
\be
T = \frac{\kappa}{2\pi} = \frac{f^\prime(r_h)}{4\pi\big(r_h^2 +n^2\big)}
 = \frac{1}{4\pi{r_h}}\Big(1 -\frac{q^2}{r_h^2 +n^2}\Big) \, ,
\ee
in which a prime represents the partial derivative with respective to its variable.

Secondly, \textcolor{black}{the total electric charge distribution over a two-dimensional sphere
with a finite radius $r$ is given by the Gauss' integral}
\bea
Q(r) = \frac{-1}{4\pi}\int_{S^2} {^\star}F = q\frac{r^2-n^2}{r^2+n^2} \, , \nn
\eea
\textcolor{black}{which clearly shows it is radial-dependently distributed and should be on the Misner
string singularities \cite{PRD12-3019}. So the electric charge on the event horizon is}
\be
Q_h = q\frac{r_h^2-n^2}{r_h^2+n^2} \, .
\ee

The corresponding electrostatic potential at the event horizon simply reads
\be
\Phi = (\mbA_{\mu}\chi^{\mu})|_{r=r_h} = \frac{qr_h}{r_h^2 +n^2}\, ,
\ee
where $\chi = \p_t$ is the timelike Killing vector normal to the event horizon.

In the Lorentzian RN-NUT spacetime, there are additional Killing horizons (north/south pole
axes) connected to the Misner strings \textcolor{black}{with the associated Misner potential
being}
\be\label{psi}
\psi = \frac{1}{8\pi n} \, .
\ee

\textcolor{black}{In the language of exterior differential forms, the Hodge dual two-form
corresponding to the generalized Komar superpotential (\ref{gKsp}) can be obtained as
\bea
{^\star}\Xi[\chi] &=& \frac{2n}{\big(r^2+n^2\big)^2}\Big(-f +\frac{4q^2r^2}{r^2+n^2}\Big)
 dr\wedge (dt +2n\cos\theta{}d\phi) \nn \\
&& -\bigg[f^{\prime} -\frac{2rf}{r^2+n^2} +\frac{2q^2r\big(r^2-n^2\big)}{
 \big(r^2+n^2\big)^2}\bigg]\sin\theta{}d\theta \wedge{}d\phi \, , \qquad
\label{mKp}
\eea
for the timelike Killing vector $\chi = \p_t$.}

\textcolor{black}{Using the above generalized Komar superpotential two-form (\ref{mKp}) to
replace the ordinary Komar one and following the same pattern of the ($\psi-\cN$)-pair
formalism as did in Refs. \cite{CQG36-194001,PRD100-104016} (namely, one deliberately
separates the integral into three parts: the spatial infinity, the horizon, and two
Misner string tubes), one can derive the integral Bekenstein-Smarr-like mass formula}
\be
M = 2TS +\Phi{}Q_h +2\psi\cN \, ,
\ee
\textcolor{black}{and then verify that the differential first law can also be satisfied}:
\be
dM = TdS +\Phi{}dQ_h +\psi{}d\cN \, .
\ee

\textcolor{black}{In the derivation of the Smarr-like formula, one can define the conserved
mass as
\be
M = \frac{-1}{8\pi}\int_{S^2_{\infty}} {^\star}\Xi[\chi]
 = \frac{-1}{4}\int_0^{\pi}d\theta (\sqrt{-g}\Xi^{tr})\big|_{r\to\infty} = m \, ,
\ee
which exactly coincides with the Komar mass calculated via the usual Komar integral at infinity,
since the well fall-off asymptotic behavior of the Maxwell field. In the computation, we has
used the determinant $\sqrt{-g} = \big(r^2+n^2\big)\sin\theta$. On the hand hand, we have instead
\be
\frac{-1}{8\pi}\int_{S^2_h} {^\star}\Xi[\chi] = 2TS +\Phi{}Q_h \, .
\ee}

\textcolor{black}{The last thermodynamic quantity of the Misner charge can be easily determined
via the above Bekenstein-Smarr-like mass formula as}
\be
\cN = \frac{4\pi n^3}{r_h}\bigg[-1 +q^2\frac{3r_h^2 +n^2}{
 \big(r_h^2 +n^2\big)^2}\bigg] \, , \label{mN}
\ee
\textcolor{black}{which is non-globally conserved. Alternately, it can also be evaluated via the
Misner tubes integral
\bea
\cN &=& \frac{n}{2}\int_{(T_+ -T_-)} {^\star}\Xi[\chi] = \pi{}n\int_{r_h}^{\infty}dr
 (\sqrt{-g}\Xi^{t\theta})\big|_{\theta=0}^{\theta=\pi} \nn \\
 &=& -8\pi{}n^3\frac{\big(r^2+n^2\big)(r_h -m) -q^2r_h}{\big(r_h^2+n^2\big)^2} \, , \nn
\eea
which reproduces the above expression (\ref{mN}) after using the identity $m = \big(r_h^2
 -n^2 +q^2\big)/(2r_h)$.}

\textcolor{black}{It can be further identified that all the above thermodynamic quantities
are related to the Gibbs free energy of the four-dimensional Lorentzian RN-NUT spacetime
\cite{JHEP0719119}
\be
G = M -TS -\psi\cN -\Phi{}Q_h \, ,
\ee
whose expression can be obtained} via a Wick-rotated back procedure from the Euclidean
action of the Euclidean RN-NUT spacetime:
\be\label{Euact1}
I_E = \frac{1}{16\pi}\int_M d^4x \sqrt{g}\big(R -F^2\big)
 +\frac{1}{8\pi}\int_{\p{}M} d^3x \sqrt{h}(K -K_0) \, ,
\ee
where $h$ is the determinant of the induced metric $h_{ij}$, $K$ is the trace of the extrinsic
curvature tensor defined on the boundary with this induced metric, and $K_0$ is the subtracted
one of the massless uncharged Taub-NUT solution as the reference background. \textcolor{black}{The
computation of the Euclidean action integral yields the following expression for the Gibbs free
energy}
\be
G = \frac{I_E}{\beta} = \frac{m}{2} -q^2\frac{r_h\big(r_h^2 -n^2\big)}{2\big(r_h^2 +n^2\big)^2}
\textcolor{black}{= \frac{1}{2}(M -\Phi{}Q_h)} \, ,
\ee
where $\beta = 1/T$ is the interval of the time coordinate.

\textcolor{black}{By the way, it should be noted that the above results are completely consistent
with those given in Ref. \cite{SCPMA64-260411} without any ``derivation".}

%%%%%%%%%%%%%%%%%%%%%%%%%%%%%%%
\subsection{Topological number}
%%%%%%%%%%%%%%%%%%%%%%%%%%%%%%%

Next, we will investigate the topological number of the four-dimensional Lorentzian RN-NUT
spacetime. We note that the Helmholtz free energy simply reads
\be\label{FERNNUT}
F = G +\Phi{}Q_h = M -TS -\psi\cN \, .
\ee
Replacing $T$ with $1/\tau$ in Eq. (\ref{FERNNUT}) and using $m = \big(r_h^2 -n^2 +q^2\big)
/(2r_h)$, then the generalized off-shell Helmholtz free energy is
\be
\cF = \frac{r_h}{2} -\frac{\pi\big(r_h^2 +n^2\big)}{\tau}
 +q^2\frac{r_h\big(r_h^2 -n^2\big)}{2\big(r_h^2 +n^2\big)^2} \, .
\ee
Adopting the definition of Eq. (\ref{vector}), the components of the vector $\phi$ can be
easily calculated as follows:
\bea
\phi^{r_h} &=& \frac{1}{2} -\frac{2\pi r_h}{\tau}
 -q^2\frac{r_h^4 -6n^2r_h^2 +n^4}{2\big(r_h^2 +n^2\big)^3} \, , \\
\phi^{\Theta} &=& -\cot\Theta\csc\Theta \, . \nn
\eea
By solving the equation: $\phi^{r_h} = 0$, one can arrive at a curve on the $r_h-\tau$ plane.
For the four-dimensional RN-NUT spacetime, one can obtain
\be\label{tauRN}
\tau = \frac{4\pi{}r_h\big(r_h^2 +n^2\big)^3}{\big(r_h^2 +n^2\big)^3
 -q^2\big(r_h^4 -6n^2r_h^2 +n^4\big)} \, .
\ee
We point out that Eq. (\ref{tauRN}) consistently reduces to the one obtained in the case of the
four-dimensional RN black hole \cite{PRL129-191101} when the NUT charge parameter $n$ is turned
off. Note that the generation point satisfies the constraint conditions:
\be
\frac{\p\tau}{\p{}r_h} = 0 \, , \qquad \frac{\p^2\tau}{\p{}r_h^2} > 0 \, .
\ee

%%%%%%%%%%%%%%%%%%%%%%%%%%%%%%%%%%%%%%%%%%%%%%%%%%%%
\begin{figure}[h]
\centering
\includegraphics[width=0.4\textwidth]{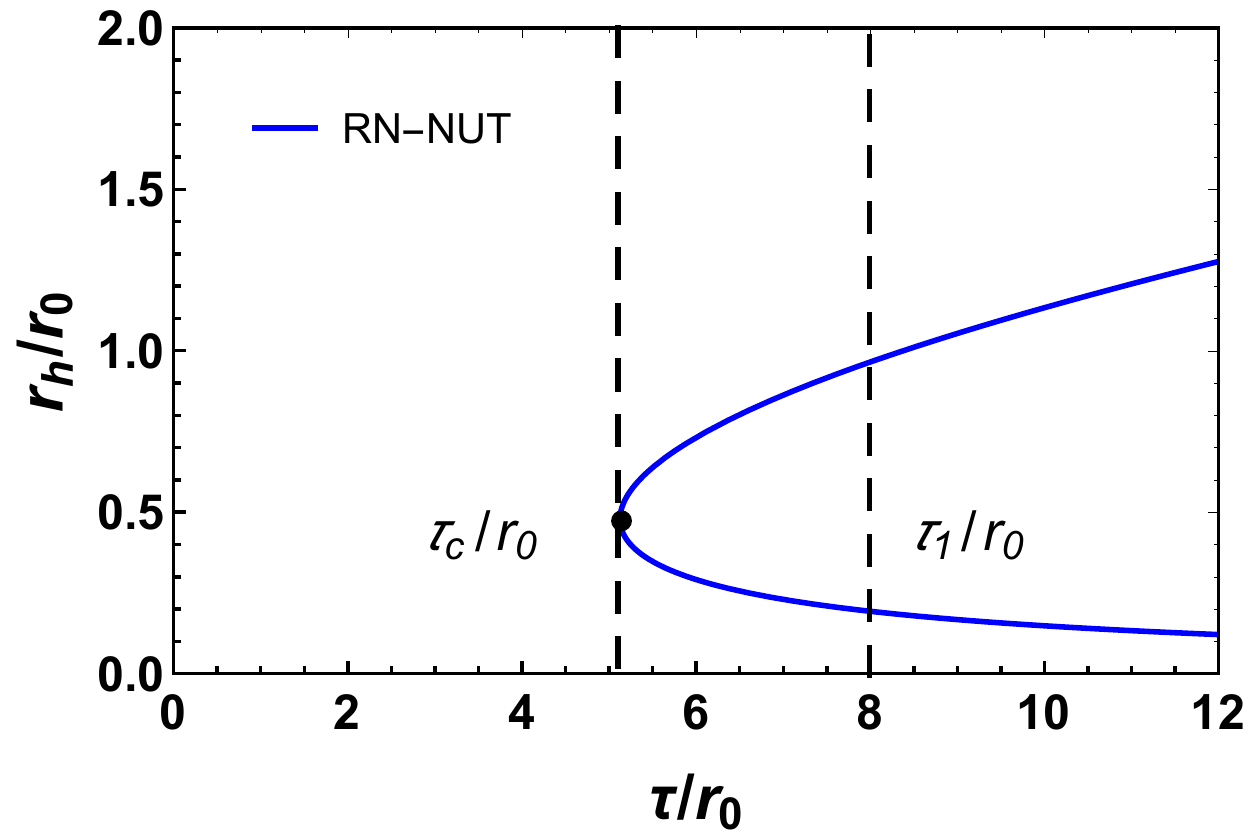}
\caption{Zero points of the vector $\phi^{r_h}$ shown in the $r_h-\tau$ plane
with $n/r_0=1$ and $q/r_0=1$. The generation point for the RN-NUT spacetime is
represented by the black dot with $\tau_c$. At $\tau = \tau_1$, there are two
RN-NUT spacetimes. \label{4dRNNUT}}
\end{figure}
%%%%%%%%%%%%%%%%%%%%%%%%%%%%%%%%%%%%%%%%%%%%%%%%%%%%%

\begin{figure}[h]
\centering
\includegraphics[width=0.4\textwidth]{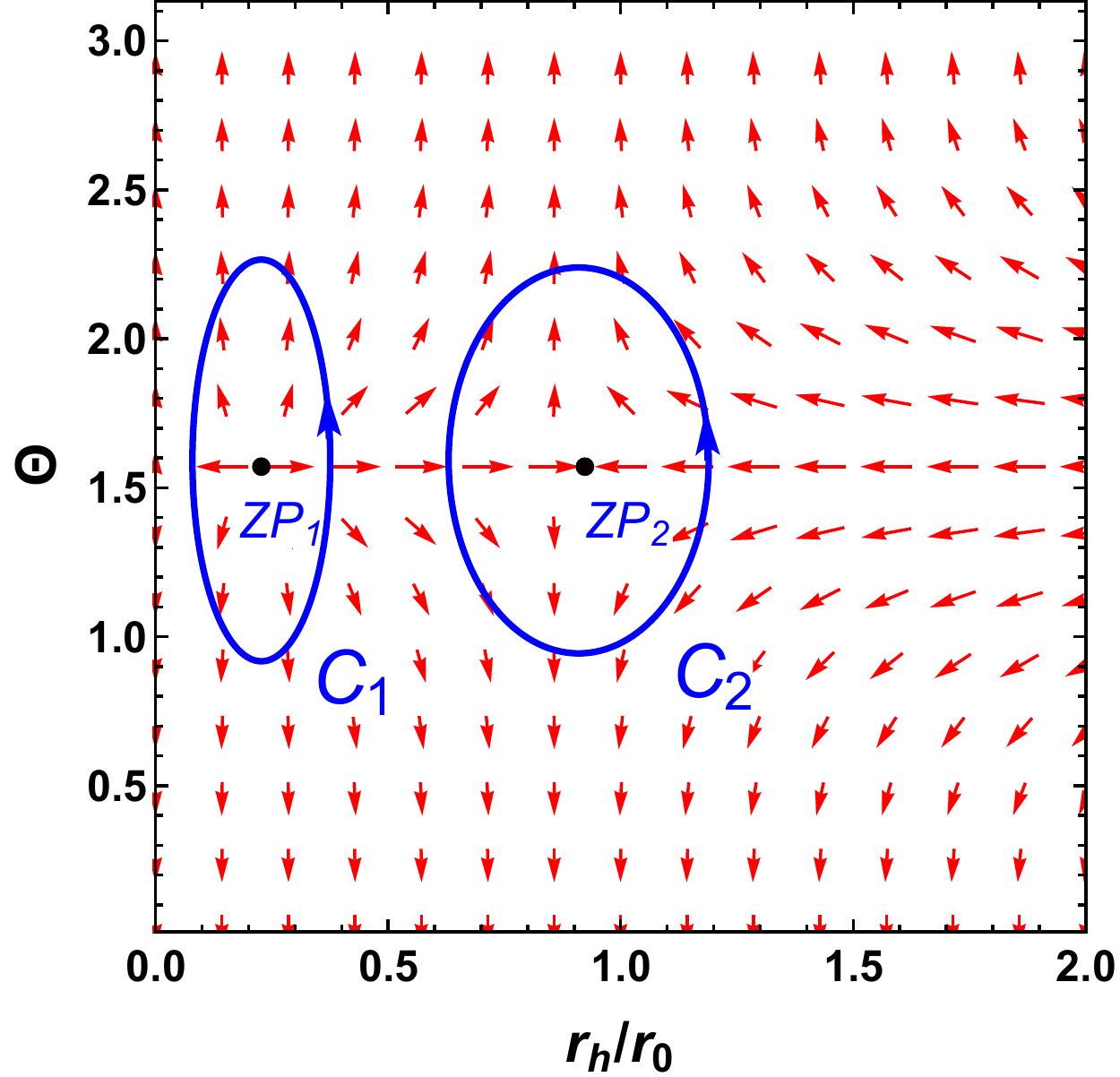}
\caption{The red arrows represent the unit vector field $n$ on a portion of the $r_h-\Theta$
plane for the RN-NUT spacetime with $n/r_0=1$, $q/r_0=1$ and $\tau/r_0 = 7$. The zero points
(ZPs) marked with black dots are at $(r_h/r_0, \Theta) = (0.23,\pi/2)$, and $(0.86,\pi/2)$,
respectively. The blue contours $C_i$ are closed loops enclosing the zero points.
\label{RNNUT4d}}
\end{figure}

Taking $q/r_0 = 1$ and $n/r_0 = 1$ for the four-dimensional Lorentzian RN-NUT spacetime, we plot
in Fig. \ref{4dRNNUT} and Fig. \ref{RNNUT4d}, respectively, for the zero points of the component
$\phi^{r_h}$, and for the unit vector field $n$ on a portion of the $\Theta -r_h$ plane with
$\tau = 7r_0$ in which $r_0$ is an arbitrary length scale set by the size of a cavity enclosing
the RN-NUT spacetime. From Fig. \ref{4dRNNUT}, one generation point can be found at $\tau/r_0 =
\tau_c/r_0 = 5.13$. It is clear that the RN-NUT spacetime behaves like the RN black hole, showing
that the NUT charge parameter appears to have no impact on the thermodynamic topological
classification for the static charged asymptotically locally flat spacetime. Consequently,
it would be fascinating to learn more about the relationship between geometric topology and
thermodynamic topology: for example, it would be very interesting to investigate the topological
number of ultraspinning black holes \cite{PRD89-084007,PRL115-031101,JHEP0114127,PRD103-104020,
PRD101-024057,PRD102-044007,PRD103-044014, JHEP1121031} and their usual counterparts.

%%%%%%%%%%%%%%%%%%%%%%%%%%%%%%%%%%%%%%%%%%%%%%%%%%

In Fig. \ref{RNNUT4d}, the zero points are located at $(r_h/r_0, \Theta) = (0.23,\pi/2)$, and
$(0.86,\pi/2)$, respectively. Thus, one can read the winding numbers $w_i$ for the blue
contours $C_i$: $w_1 = 1$, $w_2 = -1$, which are similar to those of the RN black hole
\cite{PRL129-191101}. In terms of the topological global properties, one can easily obtain
the topological number $W = 0$ for the RN-NUT spacetime from Fig. \ref{RNNUT4d}, which is
also the same one as that of the RN black hole. As a result, based upon the viewpoint of the
thermodynamic topological numbers, the Lorentzian RN-NUT spacetime should be welcomed into
the black hole family. Furthermore, it can be indicated that, while the RN-NUT spacetime and
RN black hole are evidently distinguished in geometric topology, they belong to the same class
in terms of thermodynamic topology.

%%%%%%%%%%%%%%%%%%%%%%%%%%%%%%%%%%%%
\section{KN-NUT spacetime}\label{IV}
%%%%%%%%%%%%%%%%%%%%%%%%%%%%%%%%%%%%

In this section, we will focus on the case of a rotating charged Taub-NUT spacetime by considering
the four-dimensional KN-NUT solution \cite{DN,CMP10-280,JMP10-1195,JMP14-486}, whose line element
with the Misner strings symmetrically distributed along the rotation axis is written in the
Boyer-Lindquist coordinates as:
\bea\label{KNNUT}
ds^2 &=& -\frac{\Delta(r)}{\Sigma}\big[dt +\big(2n\cos\theta
 -a\sin^2\theta\big)d\varphi\big]^2 +\frac{\Sigma}{\Delta(r)}dr^2 \nn \\
&& +\Sigma d\theta^2 +\frac{\sin^2\theta}{\Sigma}\big[a{}dt
 -\big(r^2+a^2+n^2\big)d\varphi\big]^2 \, ,
\eea
where
\be
\Sigma = r^2 +(n +a\cos\theta)^2 \, , \quad \Delta(r) = r^2 -2mr -n^2 +a^2 +q^2 \, , \nn
\ee
in which $m$, $n$, $a$ and $q$ are the mass, the NUT charge, the rotation and the electric
parameters, respectively. The event horizon radius is $r_h = m +\sqrt{m^2 +n^2 -a^2 -q^2}$.
In addition, the electromagnetic gauge potential one-form is given by
\be
\mbA = \frac{qr}{\Sigma}\big[dt +\big(2n\cos\theta -a\sin^2\theta\big)d\varphi\big] \, ,
\ee
\textcolor{black}{in a gauge that its temporal component vanishes at infinity}.

%%%%%%%%%%%%%%%%%%%%%%%%%%%%%%%%%%%%%%
\subsection{Consistent thermodynamics}
%%%%%%%%%%%%%%%%%%%%%%%%%%%%%%%%%%%%%%

Now, we investigate the consistent thermodynamics of the four-dimensional Lorentzian KN-NUT
spacetime within the framework of the ($\psi-\cN$)-pair formalism. The Bekenstein-Hawking
entropy is taken as one quarter of the event horizon area:
\be
S = \frac{A}{4} = \pi\big(r_h^2 +a^2 +n^2\big) \, .
\ee
The Hawking temperature is proportional to the surface gravity $\kappa$ on the event horizon
\be
T = \frac{\kappa}{2\pi} = \frac{f^\prime(r_h)}{4\pi\big(r_h^2 +a^2 +n^2\big)}
 = \frac{r_h -m}{2\pi\big(r_h^2 +a^2 +n^2\big)} \, .
\ee
The angular velocity at the event horizon and the Misner potential are, respectively,
\be
\Omega = \frac{a}{r_h^2 +a^2 +n^2} \, , \qquad \psi = \frac{1}{8\pi n} \, .
\ee
The electric charge $Q$ on the event horizon can be computed as
\be
Q_h = \frac{-1}{4\pi}\int_{S^2_{h}}{^\star}F
 = q\frac{\big(r_h^2 +a^2\big)^2 -n^4}{\big(r_h^2 +a^2 +n^2\big)^2 -4a^2n^2} \, ,
\ee
and its corresponding electrostatic potential at the event horizon is
\be
\Phi = (\mbA_{\mu}\chi^{\mu})|_{r=r_h} = \frac{qr_h}{r_h^2 +a^2 +n^2} \, ,
\ee
where $\chi = \p_t +\Omega\p_\varphi$ is the Killing vector normal to the event horizon.

\textcolor{black}{As for the conserved mass, one can compute it just like the non-rotating case
and get
\be
M = \frac{-1}{8\pi}\int_{S^2_{\infty}} {^\star}\Xi[\p_t] = m \, ,
\ee
for the timelike Killing vector $\p_t$. One can note that it exactly coincides with the Komar
mass evaluated via the usual Komar integral at infinity.}

\textcolor{black}{One can anticipate that both the first law and the Bekenstein-Smarr mass formula
for the Lorentzian KN-NUT spacetime should read}
\bea
dM &=& TdS +\Omega{}dJ_h +\Phi{}dQ_h +\psi{}d\cN \, , \\
M &=& 2TS +2\Omega{}J_h +\Phi{}Q_h +2\psi{}\cN \, , \label{KNSmarr}
\eea
\textcolor{black}{from which one can first solve $J_h$ in terms of $\cN$ from the integral
Smarr-like formula (\ref{KNSmarr}), and then solve $\cN = \cN(r_h, q, n, a)$ from the
differential first law. After abandoning an integration constant, one can finally get
the expressions for the gravitational Misner charge and the angular momentum as follows}:
\bea
\cN &=& \frac{4\pi n^3}{r_h}\bigg\{-1 +q^2\Big[3r_h^6
 +\big(7n^2 -3a^2\big)r_h^4 \nn \\
&& +\big(5n^4 +2a^2n^2 -7a^4\big)r_h^2 -\big(a^2 -n^2\big)^3\Big] \nn \\
&&\quad\times \big[\big(r_h^2 +a^2 +n^2\big)^2 -2a^2n^2\big]^{-2} \bigg\} \, , \\
J_h &=& a\bigg\{m +\frac{n^2}{r_h} +\frac{q^2n^2}{r_h}\Big[3r_h^6
 +\big(7a^2 -3n^2\big)r_h^4 \nn \\
&& +\big(5a^4 +2a^2n^2 -7n^4\big)r_h^2 +\big(a^2 -n^2\big)^3\Big] \nn \\
&&\quad\times \big[\big(r_h^2 +a^2 +n^2\big)^2 -4a^2n^2\big]^{-2} \bigg\} \, .
\eea
\textcolor{black}{We would like to point out that the above two expressions are identical to
those of $\tilde{N}$ and $\tilde{J}$ given by Eqs. (3.15) and (3.16) in Ref. \cite{JHEP0520084}
in the case the magnetic charge parameter is turned off, namely, the asymptotic magnetic charge
vanishes. The first law and the Bekenstein-Smarr mass formula precisely correspond to the magnetic
version of the full cohomogeneity first law \cite{JHEP0520084} when the magnetic charge parameter
is set to zero.}

\textcolor{black}{On the other hand, one can utilize the generalized Komar superpotential (\ref{gKsp})
rather than the usual Komar one with respect to the Killing vector $\chi = \p_t +\Omega\p_\varphi$
and follow the same paradigm of the ($\psi-\cN$)-pair formalism as did in Ref. \cite{PLB798-134972}
to derive the integral Bekenstein-Smarr-like mass formula (\ref{KNSmarr}) and then check that the
above thermodynamic quantities simultaneously satisfy the differential first law as well. In this way,
it is facilitated to use the GRTensor II package to perform the algebraic manipulation to get the
above involved expressions of $J_h$ and $\cN$. Here, we will not repeat the ``derivation" but
just provide a simple and equivalent way to evaluate the horizon angular momentum $J_h$ by using
our generalized Koamr superpotential (\ref{gKsp}):
\be
J_h = \frac{1}{16\pi}\int_{S_h^2} {^\star}\Xi[\p_{\varphi}]
 = \frac{1}{8}\int_0^{\pi}d\theta \sqrt{-g}\Xi^{tr}[\p_{\varphi}]\big|_{r=r_h} \, ,
\ee
which reproduces the above expression after using $m = (r_h^2 -n^2 +a^2 +q^2\big)/(2r_h)$.}

\textcolor{black}{Adopting the same procedure as did in Sec. \ref{IIIA}, the calculation of the
Euclidean action integral (\ref{Euact1}) of the KN-NUT spacetime yields the Gibbs free energy}
\be
G = \frac{m}{2} -\frac{q^2r_h\big(r^2 +a^2 -n^2\big)}{2\big(r_h^2 +a^2 +n^2\big)^2 -8a^2n^2}
\textcolor{black}{= \frac{M -\Phi{}Q_h}{2}} \, ,
\ee
\textcolor{black}{which coincides with the result of Eq. (3.1) given in Ref. \cite{JHEP0520084} in the
case when the magnetic charge parameter $g = 0$ is turned off. Furthermore, it can be also identified
as}
\be\label{GEKNNUT}
G = M -TS -\psi\cN -\Omega{}J_h -\Phi{}Q_h \, .
\ee

%%%%%%%%%%%%%%%%%%%%%%%%%%%%%%%
\subsection{Topological number}
%%%%%%%%%%%%%%%%%%%%%%%%%%%%%%%

In order to obtain the thermodynamic topological number of the KN-NUT spacetime, we need to get
the expression of the generalized off-shell Helmholtz free energy in advance. The Helmholtz free
energy is given by
\be
F = G +\Omega{}J_h +\Phi{}Q_h = M -TS -\psi\cN \, .
\ee
It is a simple matter to obtain the generalized off-shell Helmholtz free energy as
\bea
\cF &=& M -\frac{S}{\tau} -\psi\cN
 = \frac{r_h^2 +a^2}{2r_h} -\frac{\pi\big(r_h^2 +a^2 +n^2\big)}{\tau} \nn \\
&& +\frac{q^2\big(r_h^2 +a^2 -n^2\big)}{2r_h\big[\big(r_h^2 +a^2 +n^2\big)^2
 -4a^2n^2\big]^2}\Big[\big(r_h^2 +a^2\big)^3 \nn \\
&& +2\big(r_h^4 +4a^2r_h^2 -a^4\big)n^2 +\big(r_h^2 +a^2\big)n^4\Big] \, .
\eea
Then, the components of the vector $\phi$ are given by
\bea
\phi^{r_h} &=& \frac{r_h^2 -a^2}{2r_h^2}
 -\frac{q^2X}{2r_h^2\big[\big(r_h^2 +a^2 +n^2\big)^2 -4a^2n^2\big]^3} \, , \nn \\
&& -\frac{2\pi{}r_h}{\tau} \\
\phi^{\Theta} &=& -\cot\Theta\csc\Theta \, , \nn
\eea
where
\bea
X &=& r_h^{12} +3\big(2a^2-n^2\big)r_h^{10} +\big(15a^4 +39a^2n^2 -14n^4\big)r_h^8 \nn \\
&& +2\big(10a^6 +41a^4n^2 -20a^2n^4 -7n^6\big)r_h^6 \nn \\
&& +3\big(a^2 -n^2\big)\big(5a^6 +15a^4n^2 +27a^2n^4 +n^6\big)r_h^4 \nn \\
&& +\big(a^2 -n^2\big)^3\big(6a^4 +3a^2n^2 -n^4\big)r_h^2 +a^2\big(a^2 -n^2\big)^5 \, . \nn
\eea
Therefore, by solving the equation: $\phi^{r_h} = 0$, one can obtain
\be\label{tauKN}
\tau = \frac{4\pi{}r_h^3\big[\big(r_h^2 +a^2+n^2\big)^2 -4a^2n^2\big]^3}{
 \big(r_h^2 -a^2\big)\big[\big(r_h^2 +a^2+n^2\big)^2 -4a^2n^2\big]^3 -q^2X}
\ee
as the zero point of the vector field $\phi^{r_h}$. We also point out that Eq. (\ref{tauKN})
consistently reduces to the one obtained in the case of the four-dimensional Kerr-Newman black
hole \cite{PRD107-024024} when the NUT charge parameter $n$ vanishes.

\begin{figure}[h]
\centering
\includegraphics[width=0.4\textwidth]{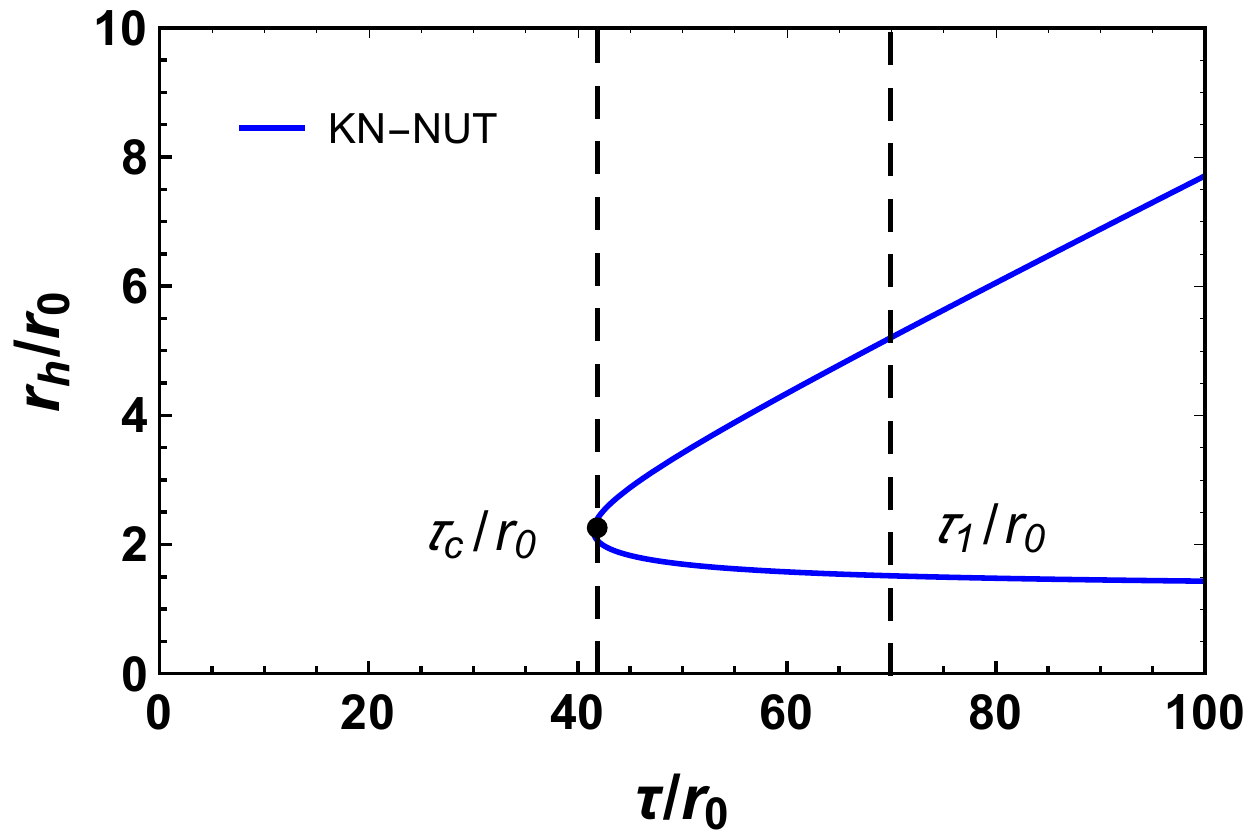}
\caption{Zero points of the vector $\phi^{r_h}$ shown in the $r_h-\tau$ plane with
$n/r_0 = 1$, $q/r_0 = 1$ and $a/r_0 = 1$. The generation point for the KN-NUT spacetime
is represented by the black dot with $\tau_c$. At $\tau = \tau_1$, there are two
KN-NUT spacetimes. Obviously, the topological number is: $W = 1 -1 =0$.
\label{4dKNNUT}}
\end{figure}

\begin{figure}[h]
\centering
\includegraphics[width=0.4\textwidth]{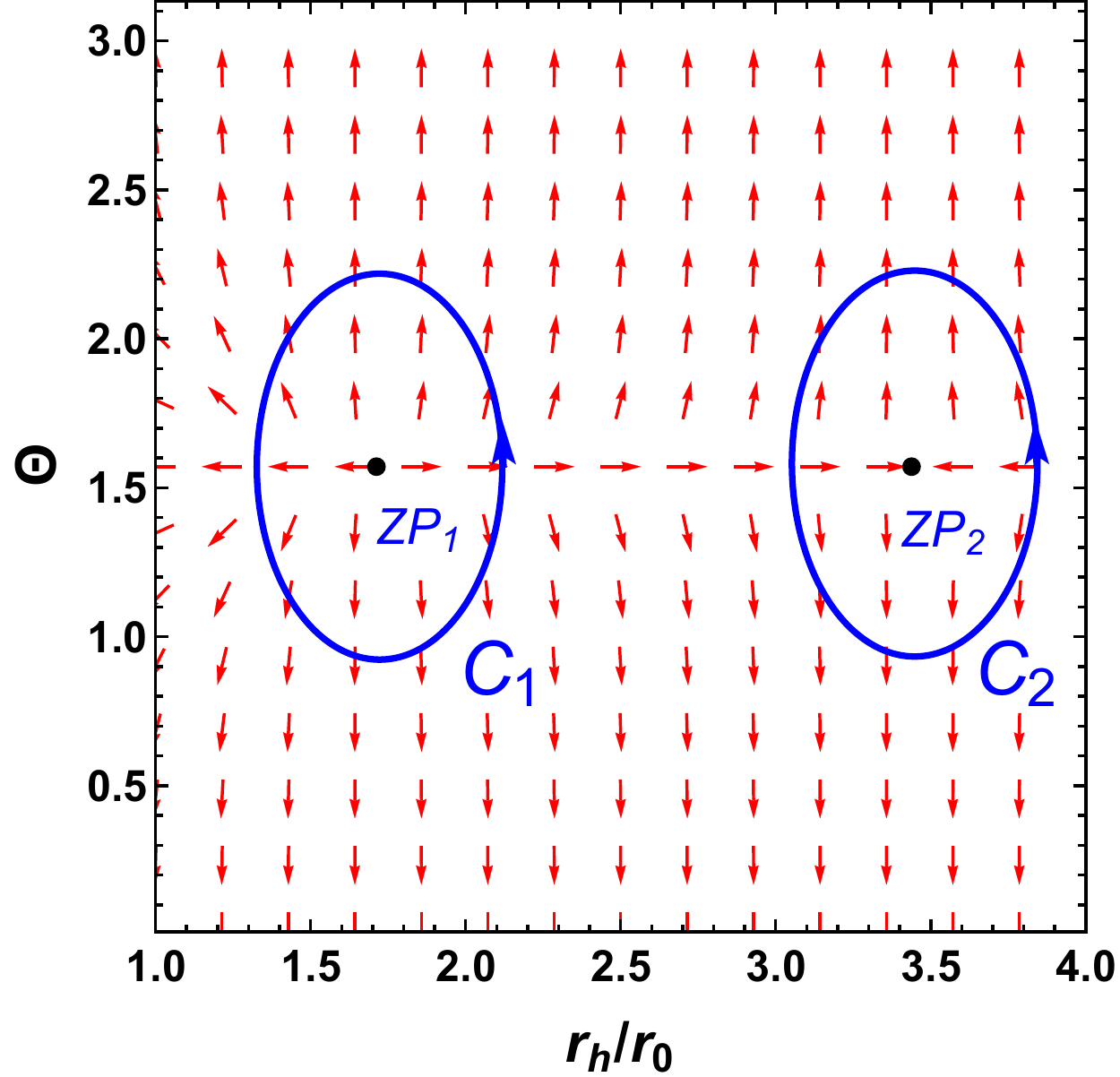}
\caption{The red arrows represent the unit vector field $n$ on a portion of the
$r_h-\Theta$ plane for the KN-NUT spacetime with $n/r_0 = 1$, $q/r_0 = 1$, $a/r_0
= 1$ and $\tau/r_0 = 50$. The zero points (ZPs) marked with black dots are at $(r_h/r_0,
\Theta) = (1.70, \pi/2)$, and $(3.41, \pi/2)$, respectively. The blue contour $C_i$
are closed loops enclosing the zero points.
\label{KNNUT4d}}
\end{figure}

Taking $n/r_0 = 1$ and $q/r_0 = 1$ as well as $n/r_0 = 1$ for the KN-NUT spacetime, we plot the
zero points of the component $\phi^{r_h}$ in Fig. \ref{4dKNNUT}, and the unit vector field $n$
on a portion of the $\Theta-r_h$ plane in Fig. \ref{KNNUT4d} with $\tau/r_0 = 50$, respectively.
In Fig. \ref{4dKNNUT}, one generation point can be found at $\tau/r_0 = \tau_{c}/r_0 = 41.49$. At
$\tau = \tau_1$, there are one thermodynamically unstable KN-NUT spacetime and one thermodynamically
stable KN-NUT spacetime, just like the Kerr-Newman black hole \cite{PRD107-024024}. In Fig.
\ref{KNNUT4d}, one can observe that the zero points are located at $(r_h/r_0, \Theta) = (1.70,
\pi/2)$, and $(3.41, \pi/2)$, respectively. Based upon the local property of the zero points,
we can obtain the topological number: $W = 1 -1 = 0$ for the KN-NUT spacetime, which is the same
one as that of the Kerr-Newman black hole \cite{PRD107-024024}. Therefore, the four-dimensional
KN-NUT spacetime should be present in the large family of black holes. Additionally, it can be
concluded that even though the KN-NUT spacetime and Kerr-Newman black hole have undoubtedly
distinct geometric topologies, they are the same type from the viewpoint of the thermodynamic
topology, just like the RN-NUT spacetime and RN black hole, which have been addressed in Sec.
\ref{III} and Ref. \cite{PRL129-191101}, respectively.

%%%%%%%%%%%%%%%%%%%%%%%%%%%%%%%%%%%%%%%%%%%%%%%%%%%%
%%%%%%%%%%%%%%%%%%%%%%%%%%%%%%%%%%%%%%%%%%%%%%%%%%%%%

%%%%%%%%%%%%%%%%%%%%%%%%%%%%%%%%%%%%%%%%%%%%%%%%%%

%%%%%%%%%%%%%%%%%%%%%%%%%%%%%%%%%%%%%%%%%%%
\section{RN-NUT-AdS$_4$ spacetime}\label{V}
%%%%%%%%%%%%%%%%%%%%%%%%%%%%%%%%%%%%%%%%%%%

In this section, we turn to explore the Lorentzian charged Taub-NUT spacetime with an negative
cosmological constant, namely, the Lorentzian RN-NUT-AdS$_4$ spacetime, whose metric and Abelian
gauge potential are still given by Eqs. (\ref{RNNUT})-(\ref{Abel}), but now $f(r) = r^2 -2mr
-n^2 +q^2 +\big(r^4 +6n^2r^2 -3n^4\big)/l^2$, in which the AdS radius $l$ is related to the
thermodynamic pressure $P = 3/\big(8\pi{}l^2\big)$ of the four-dimensional AdS black hole
\cite{CPL23-1096,CQG26-195011,PRD84-024037}. \textcolor{black}{One can show that the generalized
Komar superpotential now obey an identity: $\nabla_b\Xi^{ab} = -6\xi^a/l^2$ in the present case.}

%%%%%%%%%%%%%%%%%%%%%%%%%%%%%%%%%%%%%%
\subsection{Consistent thermodynamics}
%%%%%%%%%%%%%%%%%%%%%%%%%%%%%%%%%%%%%%

We now investigate the thermodynamical properties within the ($\psi-\cN$)-pair formalism of the
four-dimensional Lorentzian RN-NUT-AdS spacetime. \textcolor{black}{Since we are extending the
results already appeared in the subsection \ref{IIIA}, so we will mainly collect the needed
expressions and just outline the different aspects. For the event horizon $r_h$, which is the
location of the largest root of the radial function: $f(r_h) = 0$, the Bekenstein-Hawking entropy
is}
\be
S = \pi\big(r_h^2 +n^2\big) \, ,
\ee
while the Hawking temperature has a different expression
\be
T = \frac{f^\prime(r_h)}{4\pi\big(r_h^2 +n^2\big)}
= \frac{1}{4\pi{r_h}}\Big(1 -\frac{q^2}{r_h^2 +n^2}
 +3\frac{r_h^2 +n^2}{l^2}\Big) \, .
\ee
On the event horizon, the electric charge and and its corresponding electrostatic potential
\be
Q_h = q\frac{r_h^2 -n^2}{r_h^2 +n^2} \, , \qquad
\Phi = \frac{qr_h}{r_h^2 +n^2} \, .
\ee

The conformal mass can be evaluated as
\be
M = m \, ,
\ee
which is associated with the timelike Killing vector: $\chi = \p_t$. The Misner potential is
\be
\psi = \frac{1}{8\pi{}n} \, .
\ee
\textcolor{black}{Using the metric determinant: $\sqrt{-g} = \big(r^2+n^2\big)\sin\theta$, one
can define the thermodynamic volume}
\be
V = \textcolor{black}{2\pi\int_0^{\theta}d\theta\int_0^{r_h}\sqrt{-g}{}dr}
 = \frac{4}{3}\pi r_h\big(r_h^2 +3n^2\big) \, ,
\ee
which is conjugate to the pressure: $P = 3/\big(8\pi{}l^2\big)$.

Within the framework of the extended phase space, \textcolor{black}{one can substitute the above
thermodynamical quantities into the Bekenstein-Smarr mass formula}
\be
M = 2TS +\Phi{}Q_h +2\psi{}\cN +2VP \, ,
\ee
\textcolor{black}{and use the identity: $m = (r_h^2 -n^2 +q^2)/(2r_h) +\big(r_h^4 +6n^2r_h^2
-3n^4\big)/(2l^2r_h)$ to acquire the expression of the gravitational Misner charge}:

\be\label{gmc}
\cN = \frac{4\pi{}n^3}{r_h}\Big[-1 +3\frac{r_h^2 -n^2}{l^2}
 +q^2\frac{3r_h^2 +n^2}{\big(r_h^2 +n^2\big)^2} \Big] \, .
\ee
\textcolor{black}{Then one can verify that they also completely satisfy the first law}:
\be
dM = TdS +\Phi{}dQ_h +\psi{}d\cN +VdP \, .
\ee

\textcolor{black}{We would like to point out that the mass formulae presented here exactly
correspond to the magnetic-type first law and Smarr-like mass formula of the unconstrained
$\psi-\cN$ pair formalism of the consistent thermodynamics of the dyonic RN-NUT-AdS$_4$
spacetimes \cite{PRD100-104016,JHEP0719119} when the asymptotic magnetic charge is turned
off. In particular, the above expression for the Misner charge (\ref{gmc}) coincides with
that of $N^{(2)}$ explicitly given by Eq. (57) in Ref. \cite{PRD100-104016} after setting
the magnetic charge parameter to zero.}

\textcolor{black}{One can also follow the same steps as did in Refs. \cite{CQG36-194001,PRD100-104016}
to derive the above Smarr-like formula. To do so, in addition to use the generalized Komar
superpotential two-form (\ref{mKp}), one must also introduce a dual Killing co-potential
$^\star\omega$ to cancel the divergence at infinity. We shall not repeat this algebraic excise
here. Instead, there is another simple way to regulate the divergence by making a subtraction
from the massless pure NUT-charged background.}

One can obtain the Gibbs free energy \cite{JHEP0719119}
\bea\label{GERNAdS}
G &=& \frac{m}{2} -\frac{q^2r_h\big(r_h^2 -n^2\big)}{2\big(r_h^2 +n^2\big)^2}
 -\frac{r_h\big(r_h^2 +3n^2\big)}{2l^2} \nn \\
&=& \textcolor{black}{\frac{M -\Phi{}Q_h}{2} -VP} \, ,
\eea
which coincides with those computed via the Euclidean action integral, namely $G = I/\beta$. In
order to obtain this result, one can calculate the Euclidean action \cite{JHEP0719119,JHEP0321039}
for the Euclidean spacetime
\bea
I_E &=& \frac{1}{16\pi}\int_M d^4x \sqrt{g}\Big(R +\frac{6}{l^2} -F^2\Big) \nn \\
&& +\frac{1}{8\pi}\int_{\p M} d^3x \sqrt{h}\Big[K -\frac{2}{l} -\frac{l}{2}\mathcal{R}(h)\Big] \, ,
\eea
where $K$ and $\mathcal{R}(h)$ are the extrinsic curvature and Ricci scalar of the boundary metric
$h_{\mu\nu}$, respectively. In order to remove the divergence, the action includes, in addition to
the ordinary Einstein-Hilbert term, the Gibbons-Hawking boundary term and the corresponding AdS
boundary counterterms \cite{PRD60-104001,PRD60-104026,PRD60-104047,CMP208-413,CMP217-595}.
\textcolor{black}{Note that the Gibbs free energy (\ref{GERNAdS}) should also be identified with}
\be
G = M -TS -\psi\cN -\Phi{}Q_h \, .
\ee

%%%%%%%%%%%%%%%%%%%%%%%%%%%%%%%
\subsection{Topological number}
%%%%%%%%%%%%%%%%%%%%%%%%%%%%%%%

In the following, we will investigate the topological number of the four-dimensional Lorentzian
RN-NUT-AdS spacetime. The Helmholtz free energy simply reads
\be
F = G +\Phi{}Q_h = M -TS -\psi\cN \, .
\ee
Replacing $T$ with $1/\tau$ and substituting $l^2 = 3/(8\pi{}P)$, then the generalized off-shell
Helmholtz free energy is given by
\be
\cF = \frac{r_h}{2} -\frac{\pi\big(r_h^2 +n^2\big)}{\tau}
 +\frac{q^2r_h\big(r_h^2 -n^2\big)}{2\big(r_h^2 +n^2\big)^2}
 +\frac{4\pi{}P}{3}r_h\big(r_h^2 +3n^2\big) \, .
\ee
Thus, the components of the vector $\phi$ are obtained as follows:
\bea
\phi^{r_h} &=& \frac{1}{2} -\frac{2\pi{}r_h}{\tau}
 -q^2\frac{r_h^4 -6n^2r_h^2 +n^4}{2\big(r_h^2 +n^2\big)^3} \nn \\
&& +4\pi{}P\big(r_h^2 +n^2\big) \, ,  \\
\phi^{\Theta} &=& -\cot\Theta\csc\Theta \, , \nn
\eea
from which one can obtain the zero point of the vector field $\phi^{r_h}$ as
\be\label{tauRNNUTAdS}
\tau = \frac{4\pi{}r_h\big(r_h^2 +n^2\big)^3}{8\pi{}P\big(r_h^2 +n^2\big)^4
 +\big(r_h^2 +n^2\big)^3 -q^2\big(r_h^4 -6n^2r_h^2 +n^4\big)} \, ,
\ee
which consistently reduces to the one obtained in the four-dimensional RN-AdS$_4$ black hole case
\cite{PRL129-191101} when the NUT charge parameter $n$ is turned off. We also point out that the
annihilation point satisfies the constraint conditions:
\be
\frac{\p\tau}{\p{}r_h} = 0 \, , \qquad \frac{\p^2\tau}{\p{}r_h^2} < 0 \, .
\ee

\begin{figure}[h]
\centering
\includegraphics[width=0.4\textwidth]{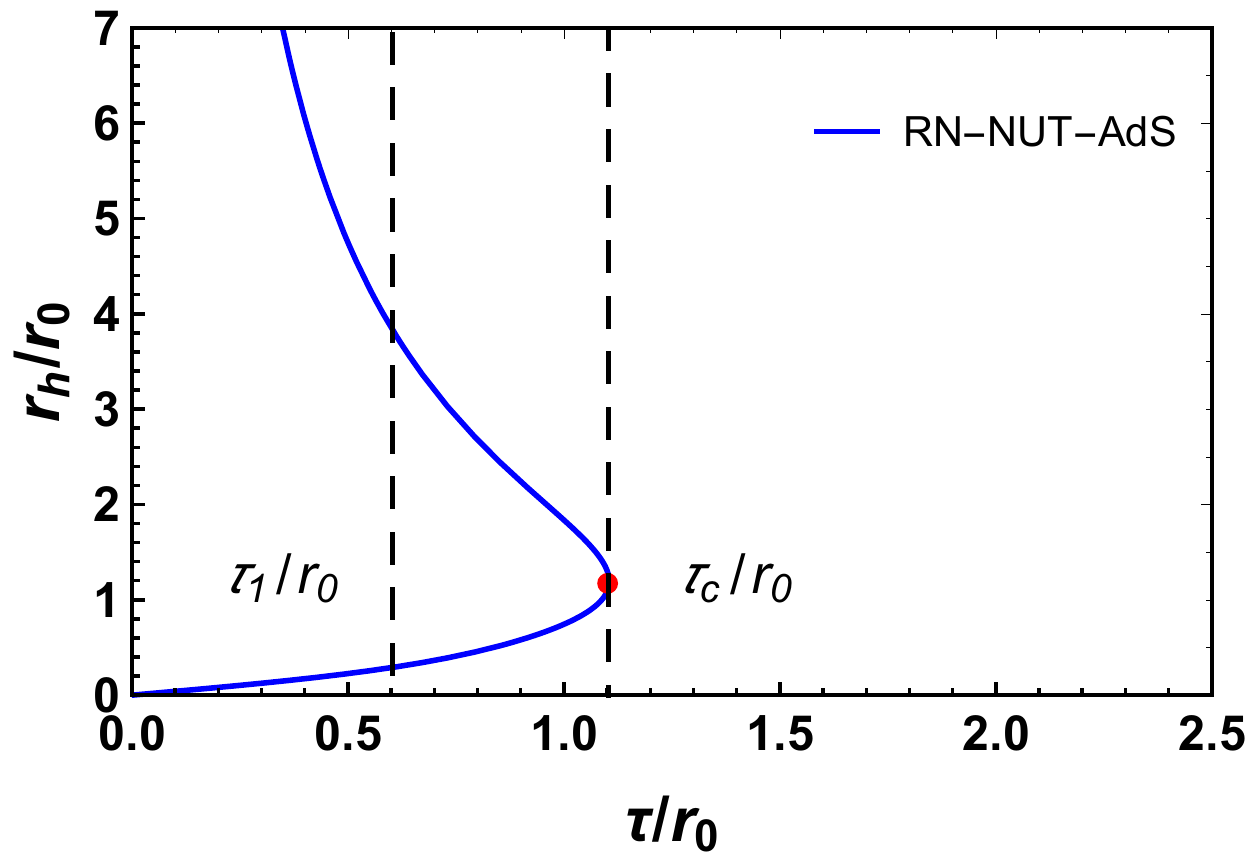}
\caption{Zero points of the vector $\phi^{r_h}$ shown on the $r_h-\tau$ plane with
$q/r_0 = 1$, $n/r_0 = 1$, and $Pr_0^2 = 0.2$ for the RN-NUT-AdS$_4$ spacetime. The
annihilation point for this spacetime is represented by the red dot with $\tau_c$.
There are two RN-NUT-AdS$_4$ spacetimes when $\tau = \tau_1$. Clearly, the topological
number is: $W = -1 +1 = 0$.
\label{4dRNNUTAdS}}
\end{figure}

\begin{figure}[h]
\centering
\includegraphics[width=0.4\textwidth]{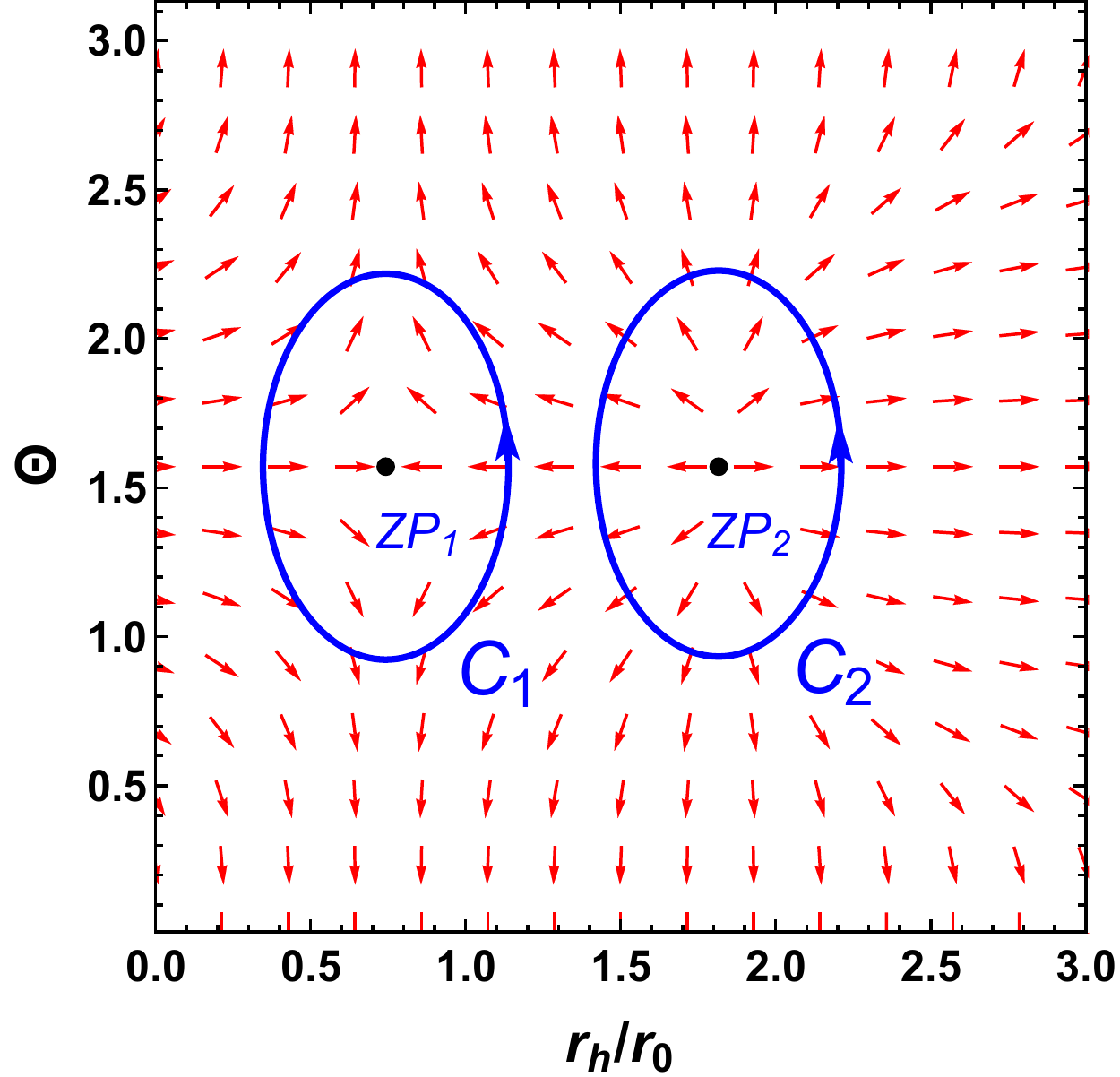}
\caption{The red arrows represent the unit vector field $n$ on a portion of the
$r_h-\Theta$ plane with $q/r_0 = 1$, $n/r_0 = 1$, $Pr_0^2 = 0.2$ and $\tau/r_0 = 1$
for the RN-NUT-AdS$_4$ spacetime. The zero points (ZPs) marked with black dots are
at $(r_h/r_0, \Theta) = (0.74, \pi/2)$, $(1.84, \pi/2)$ for ZP$_1$ and ZP$_2$,
respectively. The blue contours $C_i$ are closed loops surrounding the zero points.
\label{RNNUTAdS4d}}
\end{figure}

Taking the pressure $Pr_0^2 = 0.2$ and the NUT charge parameter $n/r_0 = 1$ as well as the electric
charge parameter $q/r_0 = 1$ for the RN-NUT-AdS$_4$ spacetime, we plot the zero points of $\phi^{r_h}$
in the $r_h-\tau$ plane in Figs. \ref{4dRNNUTAdS} and the unit vector field $n$ on a portion of the
$\Theta-r_h$ plane with $\tau = r_0$ in Fig. \ref{RNNUTAdS4d}, respectively. In Fig. \ref{4dRNNUTAdS},
one annihilation point can be found at $\tau/r_0 = \tau_c/r_0 = 1.10$. From Fig. \ref{RNNUTAdS4d},
one can find that the zero points are located at $(r_h/r_0, \Theta) = (0.74, \pi/2)$, and $(1.84,
\pi/2)$, respectively. According to the conclusions in Ref. \cite{2305.05916}, it can be inferred
that the second-order phase transition occurs in the RN-NUT-AdS$_4$ spacetime system. Thus, it is
very interesting to explore the phase transitions of the RN-NUT-AdS$_4$ spacetime, such as the
Hawking-Page phase transitions \cite{CMP87-577} and the P-V criticality \cite{JHEP0712033} to
check the correctness of the above conjecture. In addition, for the RN-NUT-AdS$_4$ spacetime,
we observe that the topological number is: $W = 0$, and is different from that of the RN-AdS$_4$
black hole, which has: $W = 1$ \cite{PRL129-191101}, \textcolor{black}{because the third zero point
of the RN-AdS$_4$ black hole solution vanishes once the NUT charge parameter is introduced}.
Therefore, it indicates that the NUT charge parameter has a remarkable effect on the topological
number for the static charged asymptotically local AdS spacetime. As a result, at least according
to the viewpoint of the thermodynamic topological approach, the Lorentzian RN-NUT-AdS$_4$ spacetime
should be included into a member of the black hole family.

%%%%%%%%%%%%%%%%%%%%%%%%%%%%%%%%%%%%%
%%%%%%%%%%%%%%%%%%%%%%%%%%%%%%%%%%%%%%

%%%%%%%%%%%%%%%%%%%%%%%%%%%%%%%%%%%%%%%%%%%%%%%%%%%
%%%%%%%%%%%%%%%%%%%%%%%%%%%%%%%%%%%%%%%%%%%%%%%%%%%

%%%%%%%%%%%%%%%%%%%%%%%%%%%%%%%
\section{Conclusions}\label{VI}
%%%%%%%%%%%%%%%%%%%%%%%%%%%%%%%

Our results found in the present paper are now summarized in the following Table \ref{I}.
\begin{table}[h]
\caption{The topological number $W$, numbers of generation and annihilation
points for the four-dimensional Lorentzian charged Taub-NUT spacetimes.}
\resizebox{0.48\textwidth}{!}{
\begin{tabular}{c|c|c|c}
\hline\hline
Solutions & $W$ & Generation point & Annihilation point\\ \hline
RN-NUT & 0 & 1 & 0\\
KN-NUT & 0 & 1 & 0\\
RN-NUT-AdS & 0 & 0 & 1\\
\hline\hline
\end{tabular}}
\label{I}
\end{table}

In this paper, we first derive the consistent thermodynamics of the four-dimensional Lorentzian
charged RN-NUT, KN-NUT, and RN-NUT-AdS spacetimes within the framework of the ($\psi-\cN$)-pair
formalism, \textcolor{black}{which exactly correspond to the magnetic version of the full
cohomogeneity (unconstrained) first law \cite{JHEP0520084,PRD100-104016,JHEP0719119} of the
dyonic NUT-charged spactimes when the magnetic charge parameter vanishes. Then we} investigate their
topological numbers by using the uniformly modified form of the generalized off-shell Helmholtz free
energy. We found that the RN-NUT spacetime has: $W = 0$, which is the same one as that of the RN
black hole \cite{PRL129-191101}. We showed that the KN-NUT spacetime has: $W = 0$, which is identical
to that of the Kerr-Newman black hole \cite{PRD107-024024}. In addition, we also indicated that the
RN-NUT-AdS$_4$ spacetime has: $W = 0$, which is different from that of the RN-AdS$_4$ black hole
($W = 1$) \cite{PRL129-191101}. Therefore, one can conclude that although the existence of the NUT
charge parameter seems to have no impact on the topological number of the charged asymptotically locally
flat spacetimes, it has an important effect on the topological number of the charged asymptotically
locally AdS spacetime. Furthermore, it can be demonstrated that the four-dimensional RN-NUT, KN-NUT and
RN-NUT-AdS spacetimes should be treated as generic black holes from the standpoint of the thermodynamic
topological approach.

There are two promising further topics that can be pursued in the future. As mentioned above, one
intriguing topic is to explore the phase transitions of the RN-NUT-AdS$_4$ spacetime. Another one
is to extend the present work to the more general dyonic cases \cite{PRD105-124013} and higher-even
dimensional cases \cite{2209.01757,2306.00062}.

\acknowledgments

\textcolor{black}{We are greatly indebted to the anonymous referee for his/her constructive comments
to improve the presentation of this work. We also thank Professor Shuang-Qing Wu for useful
discussions.} This work is supported by the National Natural Science Foundation of China (NSFC) under
Grant No. 12205243, No. 11675130, by the Sichuan Science and Technology Program under Grant No.
2023NSFSC1347, and by the Doctoral Research Initiation Project of China West Normal University under
Grant No. 21E028.

\textcolor{black}{\textit{Notes added}.--After the submission of this paper, we became aware of a new
preprint \cite{2306.05266}, which independently presents the same result for the first law of the
KN-NUT spacetime via an alternative method.}

\end{document}